\documentclass{appolb}
\usepackage{graphicx}
\usepackage{pstricks}
\usepackage{amssymb,amsmath}
\usepackage{bbm}
\begin{document}

\newrgbcolor{orange}{1.00 0.65 0.00}
\newrgbcolor{mellow}{0.847 0.72 0.525}
\newrgbcolor{bisque}{1.0000 0.7901 0.5563}
\newrgbcolor{peach}{1.0000 0.7156 0.4846}
\newrgbcolor{azure}{0.8616 1.0000 1.0000}
\newrgbcolor{lavender}{0.7753 0.8052 0.9597}
\newrgbcolor{lightblue}{0.3779 0.7009 0.8022}
\newrgbcolor{cyan}{0.0000 1.0000 1.0000}
\newrgbcolor{lightcyan}{0.7260 1.0000 1.0000}
\newrgbcolor{aquamarine}{0.1709 1.0000 0.6678}
\newrgbcolor{palegreen}{0.2713 0.9684 0.2986}
\newrgbcolor{khaki}{0.8616 0.8052 0.2423}
\newrgbcolor{indianred}{0.5812 0.0822 0.0813}
\newrgbcolor{lightsalmon}{1.0000 0.3434 0.1702}
\newrgbcolor{pink}{1.0000 0.5343 0.6041}
\newrgbcolor{magenta}{1.0000 0.0000 1.0000}
\newrgbcolor{black}{0 0 0}
\newrgbcolor{limegreen}{0.0164 0.6223 0.0164}
\newrgbcolor{darkorange}{1.0000 0.2455 0.0000}
\newrgbcolor{firebrick}{1.0000 0.0141 0.0149}
\newrgbcolor{redII}{0.8441 0.0000 0.0000}
\newrgbcolor{skyblue}{0.3999 0.6294 0.9597}
\newrgbcolor{skyblueI}{0.4999 0.7294 0.9597}
\newrgbcolor{skyblueII}{0.5999 0.8294 0.9597}
\newrgbcolor{deepskyblue}{0.0000 0.5278 1.0000}
\newrgbcolor{lighttext}{.47 .61 .74}
\newrgbcolor{black}{0 0 0}
\newrgbcolor{grayc}{0.8 0.4 0.8}
\newrgbcolor{grayd}{1.0 0.7 0.4}
\newrgbcolor{blueIII}{0.3281 0.3281 1.0}

\def\colora{}
\def\colorb{}
\def\colorc{}
\def\colord{}
\def\colore{}
\def\colorf{}
\def\colorg{}
\def\colori{}

\def\empile#1\over#2{\mathrel{\mathop{\kern 0pt#1}\limits_{#2}}}
\newcommand{\slvarepsilon}{\raise.15ex\hbox{$/$}\kern-.53em\hbox{$\varepsilon$}}
\newcommand{\slL}{\raise.15ex\hbox{$/$}\kern-.53em\hbox{$L$}}
\newcommand{\slP}{\raise.15ex\hbox{$/$}\kern-.53em\hbox{$P$}}
\newcommand{\slD}{\raise.15ex\hbox{$/$}\kern-.53em\hbox{$D$}}
\newcommand{\slp}{\raise.1ex\hbox{$/$}\kern-.63em\hbox{$p$}}
\newcommand{\slq}{\raise.1ex\hbox{$/$}\kern-.53em\hbox{$q$}}
\newcommand{\slv}{\raise.1ex\hbox{$/$}\kern-.63em\hbox{$v$}}
\newcommand{\slR}{\raise.15ex\hbox{$/$}\kern-.53em\hbox{$R$}}
\newcommand{\slQ}{\raise.15ex\hbox{$/$}\kern-.53em\hbox{$Q$}}
\newcommand{\slK}{\raise.15ex\hbox{$/$}\kern-.53em\hbox{$K$}}
\newcommand{\slk}{\raise.15ex\hbox{$/$}\kern-.53em\hbox{$k$}}
\newcommand{\slSigma}{\raise.15ex\hbox{$/$}\kern-.53em\hbox{$\Sigma$}}
\newcommand{\slcalP}{\raise.15ex\hbox{$/$}\kern-.63em\hbox{$\cal P$}}
\newcommand{\slcalA}{\raise.15ex\hbox{$/$}\kern-.63em\hbox{$\cal A$}}
\newcommand{\slA}{\raise.15ex\hbox{$/$}\kern-.73em\hbox{$A$}}
\newcommand{\slbfA}{\raise.15ex\hbox{$/$}\kern-.73em\hbox{${\imb A}$}}
\newcommand{\slpartial}{\raise.15ex\hbox{$/$}\kern-.53em\hbox{$\partial$}}
\newcommand{\sla}{\raise.15ex\hbox{$/$}\kern-.53em\hbox{$a$}}
\newcommand{\slb}{\raise.15ex\hbox{$/$}\kern-.53em\hbox{$b$}}
\newcommand{\slc}{\raise.15ex\hbox{$/$}\kern-.53em\hbox{$c$}}
\newcommand{\slC}{\raise.15ex\hbox{$/$}\kern-.63em\hbox{$C$}}

\def\p{{\boldsymbol p}}
\def\q{{\boldsymbol q}}
\def\P{{\boldsymbol P}}
\def\l{{\boldsymbol l}}
\def\k{{\boldsymbol k}}
\def\m{{\boldsymbol m}}
\def\n{{\boldsymbol n}}
\def\x{{\boldsymbol x}}
\def\y{{\boldsymbol y}}
\def\X{{\boldsymbol X}}
\def\r{{\boldsymbol r}}
\def\s{{\boldsymbol s}}
\def\u{{\boldsymbol u}}
\def\v{{\boldsymbol v}}
\def\w{{\boldsymbol w}}
\def\z{{\boldsymbol z}}
\def\b{{\boldsymbol b}}
\def\a{{\boldsymbol a}}
\def\A{{\boldsymbol A}}
\def\E{{\boldsymbol E}}
\def\B{{\boldsymbol B}}
\def\cc{{\boldsymbol c}}

\def\bs{\boldsymbol}

\def\wt#1{\widetilde{#1}}

% \eqsec  % uncomment this line to get equations numbered by (sec.num)
\title{Factorization in high energy nucleus-nucleus collisions%
\thanks{Presented at ISMD 2012, Kielce, Poland.}%
% you can use '\\' to break lines
}
\author{Fran\c cois Gelis
\address{Institut de Physique Th\'eorique, CEA Saclay, 
91191 Gif sur Yvette  cedex, France}
}
\maketitle
\begin{abstract}
  In this talk, we discuss the factorization of the logarithms of
  energy in the Color Glass Condensate framework.
\end{abstract}
  
\section{Color Glass Condensate}
Heavy ion collisions at high energy are a situation in which the
colliding projectiles contain high densities of gluons, and are
possibly subject to the phenomenon of gluon
saturation~\cite{GriboLR1,MuellQ1,Lappi6}: when the gluon occupation
number approaches the inverse of the strong coupling constant
$\alpha_s$, the interactions among the gluons become strong, and
non-linear effects such as recombination become important.
\begin{figure}[htbp]
\begin{center}
\hfil
\resizebox*{4.5cm}{!}{\includegraphics{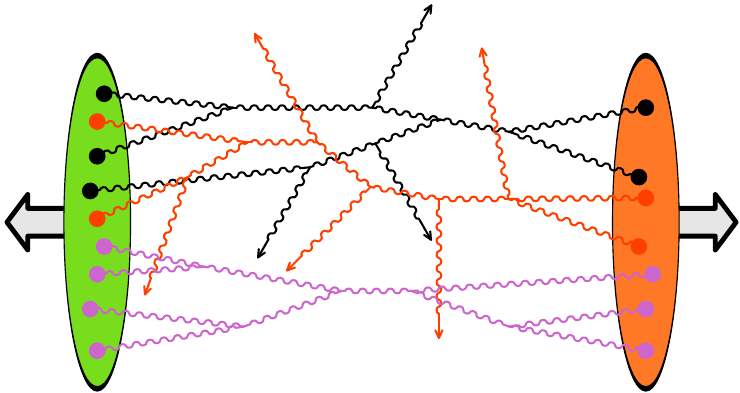}}
\hfil
\resizebox*{4.5cm}{!}{\includegraphics{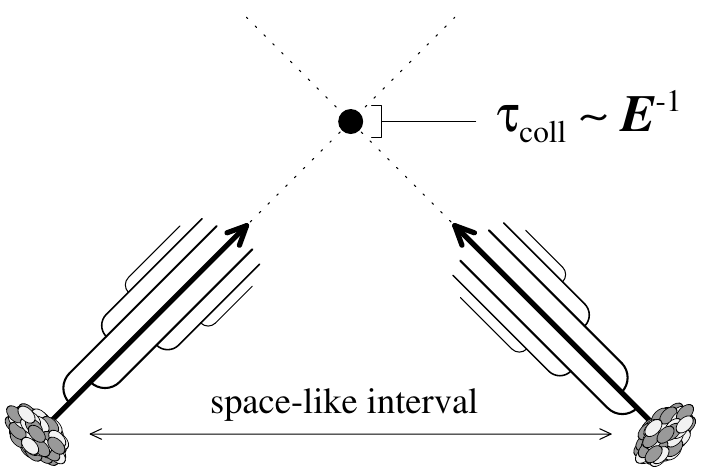}}
\hfil
\end{center}
\caption{\label{fig:cms} Left: typical graphs in the collision of two saturated projectiles. Right: Causality argument underlying factorization.}
\end{figure}

Moreover, in this regime a typical collision involves multiple gluon
interactions, as depicted in the figure \ref{fig:cms} (left). Unlike
in the dilute regime, multiple gluons from each projectile can
participate in the reaction. This leads to two complications: we a way
to organize and resum all the relevant graphs, and we need multi-gluon
distributions to describe the projectiles. Both are provided by the
Color Glass Condensate effective
theory~\cite{IancuLM3,IancuV1,GelisIJV1}, in which the incoming nuclei
are described as collections of color charges moving at the speed of
light along the light-cones.

The object that encodes this information is a probability distribution
$W[\rho]$, where $\rho(\x_\perp)$ is the color charge per unit of
transverse area of a given nucleus.  In the saturated regime, this
density is inversely proportional to the coupling, and observables at
leading order are obtained by summing an infinite series of tree
diagrams, which can be done more conveniently by solving the classical
Yang-Mills equations, in which the charge density $\rho$ plays the
role of a source term. 

Next-to-leading order corrections are made of one-loop diagrams in a
classical background field. When computing NLO corrections, the CGC
should be viewed as an effective theory with an upper cutoff $\Lambda$
on the longitudinal momentum running in the loop (see the figure
\ref{fig:cgc}, left), in order to prevent double counting the modes already
taken into account via the source $\rho$. NLO corrections in general
contain logarithms of this unphysical cutoff. For the CGC framework to
be consistent, observables should be cutoff independent in the end.
For a given observable, it is possible to make the distribution
$W[\rho]$ cutoff dependent in order to cancel the cutoff dependence
coming from the loop correction. However, for this strategy to be
useful in practice, two requirements should be satisfied~:
\begin{itemize}
\item[{\bf i}.] the same $W[\rho]$ should be able to cancel the cutoff
  dependence of a wide range of observables,
\item[{\bf ii.}] when colliding two such projectiles, each of them
  should have its own $W[\rho]$. Moreover, these distributions should
  be the same as in simpler collisions such as Deep Inelastic
  Scattering (DIS), that involve only one saturated projectile.
\end{itemize}
Under these conditions, the distribution $W[\rho]$ can be viewed as an
intrinsic property of a saturated projectile, and its cutoff
dependence reflects changes in its apparent color content due to a
change in the resolution scale.

\section{Factorization in Deep Inelastic Scattering}
\begin{figure}[htbp]
\begin{center}
\hfil
\resizebox*{6cm}{!}{\includegraphics{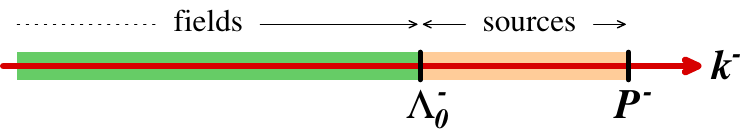}}
\hfil
\resizebox*{2.4cm}{!}{\includegraphics{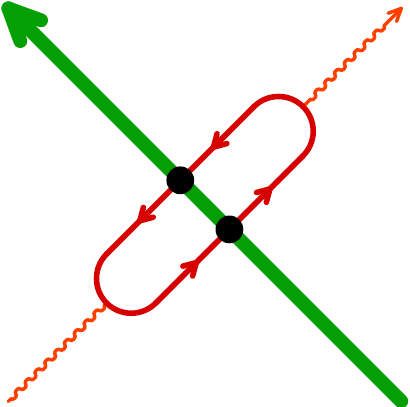}}
\hfil
\end{center}
\caption{\label{fig:cgc}Left: separation between sources and fields in
  the CGC effective theory. Right: DIS reaction at leading order.}
\end{figure}
Let us start with a reminder of the situation in inclusive DIS off a
dense nucleus. Thanks to the optical theorem, the total DIS
cross-section is related to the forward elastic scattering amplitude
of the virtual photon on the nucleus, and at LO it corresponds to the
scattering of a quark-antiquark fluctuation of the photon off the
classical color field ${\cal A}^\mu$ of the nucleus (see the figure
\ref{fig:cgc}, right)~:
\begin{eqnarray}
{\colorb {\bs T}_{_{\rm LO}}(\vec\x_\perp,\vec\y_\perp)}
&=&
1-\frac{1}{N_c}\,{\rm tr}\,({{\colord U(\vec\x_\perp)U^\dagger(\vec\y_\perp)}})
\nonumber\\
U(\vec\x_\perp)&=& {\rm P}\,\exp \,i{\colord g}\int^{1/x P^-} 
\!\!\!\!\!\! dz^+\,{\colorb{\cal A}^-(z^+,\vec\x_\perp)}
\; .
\end{eqnarray}
Note that the $q\bar{q}$ pair couples only to the sources up to the
longitudinal coordinate $z^+\lesssim (x P^-)^{-1}$. The other sources
are too slow to be seen by the dipole.

\begin{figure}[htbp]
\begin{center}
\hfil
\resizebox*{6cm}{!}{\includegraphics{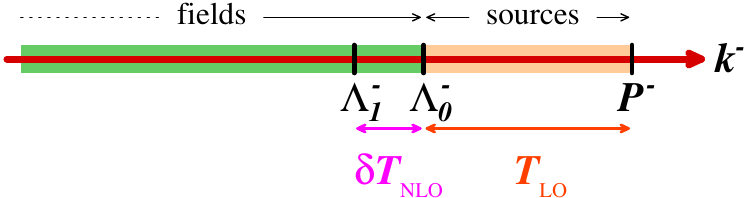}}
\hfil
\resizebox*{2.4cm}{!}{\includegraphics{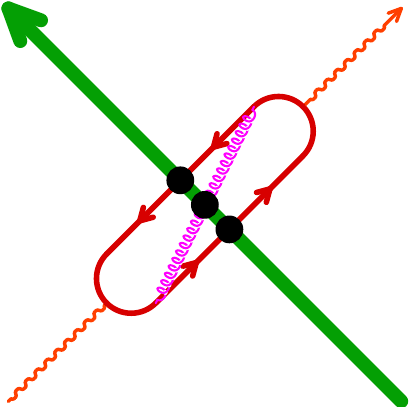}}
\hfil
\end{center}
\caption{\label{fig:dis-nlo}Left: small shift of the cutoff scale. Right: DIS reaction at next to leading order.}
\end{figure}

At NLO, one needs corrections involving a gluon, such as the one in
the right of the figure \ref{fig:dis-nlo}. The longitudinal momentum
of this gluon should be integrated only up to the cutoff $\Lambda_0^-$
of the CGC effective theory. In practice, it is more convenient to
include only longitudinal modes in a slice $\Lambda_1^-<
k^-<\Lambda_0^-$.
At leading log accuracy, the contribution of the quantum modes in that
strip is~:
\begin{equation}
  {\colorc\delta {\bs T}_{_{\rm NLO}}(\vec\x_\perp,\vec\y_\perp)}
  =
  \ln\left(\frac{\Lambda_0^-}{\Lambda_1^-}\right)\;
  {\colorb{\cal H}}\;
  {\colord{\bs T}_{_{\rm LO}}(\vec\x_\perp,\vec\y_\perp)}\; ,
\end{equation}
where ${\cal H}$ is an operator that contains functional derivatives
with respect to the source $\rho$, called the JIMWLK Hamiltonian.
These NLO corrections can be absorbed in the LO result,
\begin{equation}
\Big<{\colord{\bs T}_{_{\rm LO}}}+{\colorc\delta{\bs T}_{_{\rm NLO}}}\Big>_{\Lambda_0^-}
=
\Big<{\colord{\bs T}_{_{\rm LO}}}\Big>_{\Lambda_1^-}\; ,\quad
\big<\cdots\big>_{_\Lambda}\equiv \int[D\rho]\,W_\Lambda[\rho]\,\cdots
\end{equation}
provided one defines a new effective theory with a lower cutoff
$\Lambda_1^-$ and a modified distribution of sources
$W_{\Lambda_1^-}[\rho]$~:
\begin{equation}
W_{\Lambda_1^-}\equiv \Big[1+\ln\left(\frac{\Lambda_0^-}{\Lambda_1^-}\right)\;
    {\colorb{\cal H}}\Big]\;W_{\Lambda_0^-}\; ,\quad \mbox{i.e.}\quad
 \frac{\partial W_{\Lambda}}{\partial\ln(\Lambda)} = -{\cal H}\,W_{\Lambda}\; .
\end{equation}
By solving this equation (known as the JIMWLK equation) until the
cutoff has been lowered to the value $x P^-$, one can sequentially
resum the leading log contributions from each of the $k^-$ slices
below the original cutoff.

\section{Factorization in nucleus-nucleus collisions}
In high energy nucleus-nucleus collisions, there are now two saturated
projectiles, and it is more difficult to extract the logarithms that
arise in observables at NLO.  There is however a simple argument
explaining qualitatively the universality of the distributions
$W[\rho]$ that describe these projectiles in the CGC framework,
illustrated in the figure \ref{fig:cms} (right). The soft gluon
radiation responsible for the logarithms of the cutoff takes a long
time, much shorter than the collision time that goes like the inverse
of the collision energy. Therefore, these gluons must be emitted
before the collision, at a time where the separation between the
projectiles is space-like.  For this reason, there cannot be any
cross-talk between the distributions $W[\rho]$ that describe the two
nuclei, and they should be the same as in DIS.

\begin{figure}[htbp]
\begin{center}
\hfil
\resizebox*{2.4cm}{!}{\includegraphics{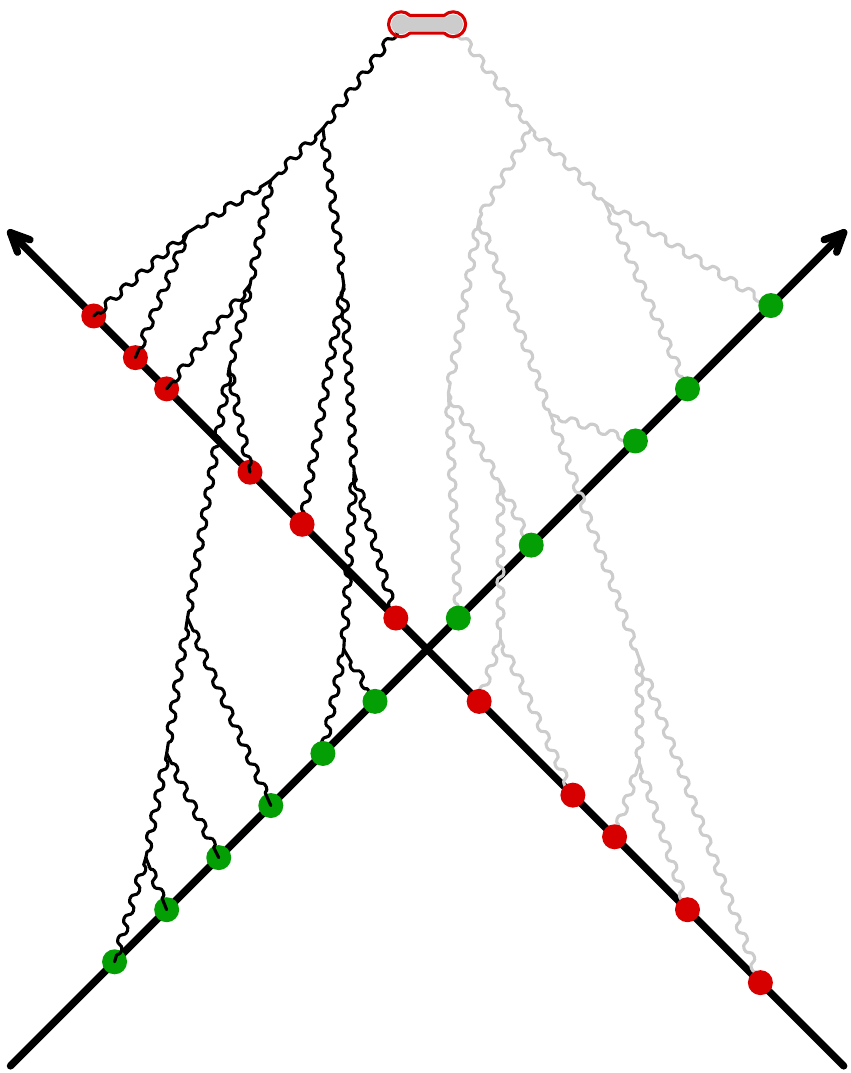}}
\hskip 5mm
\resizebox*{2.4cm}{!}{\includegraphics{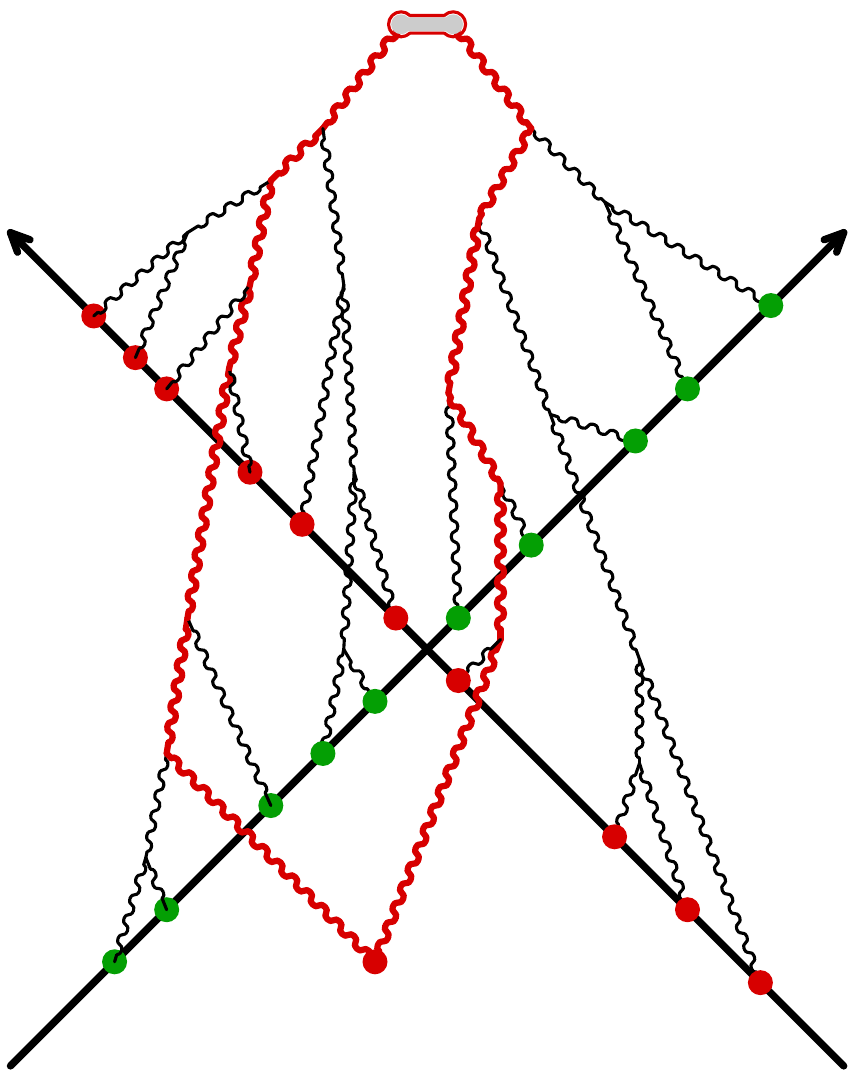}}
\hfill
\resizebox*{2.4cm}{!}{\includegraphics{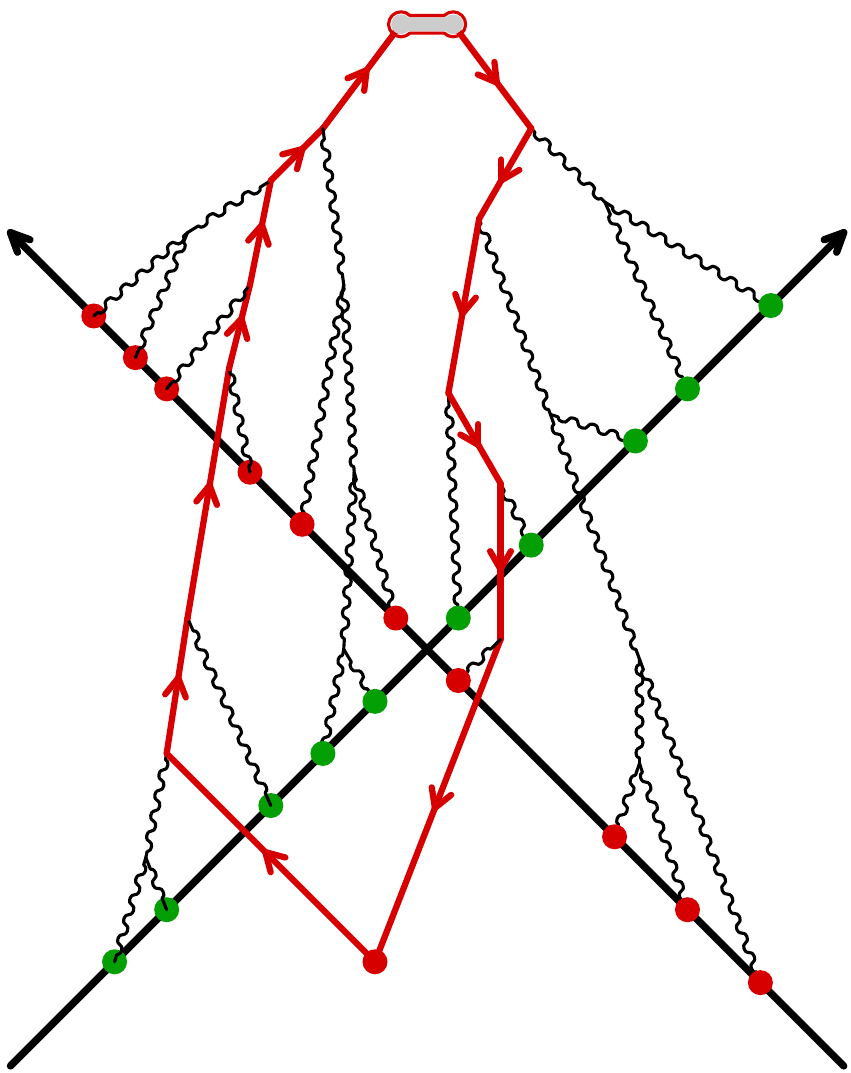}}
\hskip 5mm
\resizebox*{2.4cm}{!}{\includegraphics{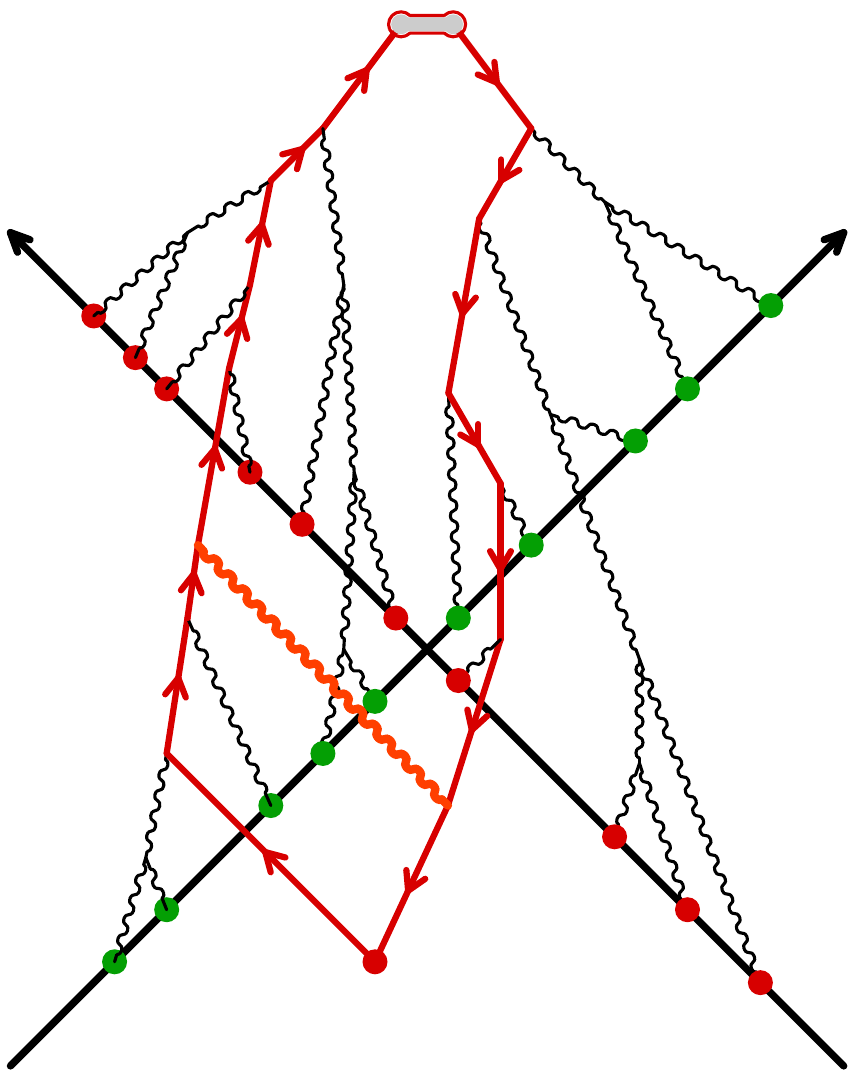}}
\hfil
\end{center}
\caption{\label{fig:gluon}Left: inclusive gluon spectrum at LO and
  NLO. Right: inclusive quark spectrum at LO and NLO.}
\end{figure}

In order to illustrate how this works in an actual calculation, let us
consider the inclusive gluon spectrum at LO and NLO, illustrated in
the two diagrams on the left of the figure \ref{fig:gluon}.
At LO, the gluon spectrum is quadratic in the retarded solution of the
classical Yang-Mills equation,
\begin{equation}
\left.\frac{dN_g}{dyd^2\vec\p_\perp}\right|_{_{\rm LO}}\propto
\int d^4x d^4y\; e^{ip\cdot(x-y)} \cdots
{\colord{\cal A}^\mu(x){\cal A}^\nu(y)}\; ,
\end{equation}
\begin{equation}
\quad{\big[{\colord{\cal D}_\mu},{\colord{\cal F}^{\mu\nu}}\big]={\colorb J_1^\nu}+{\colorb J_2^\nu}}\quad,\quad
\lim_{t\to -\infty}{\colord{\cal A}^\mu(t,\vec\x)}=0\; .
\end{equation}
At NLO, it involves one-loop graphs in the presence of a background
color field. One can formally express the NLO result as\cite{GelisLV3}
\begin{equation}
\left.{\colorb\frac{dN_g}{dyd^2\vec\p_\perp}}\right|_{_{\rm NLO}}
=
\Bigg[
 \frac{1}{2}\!\!\!\!
\int\limits_{_{\vec\u,\vec\v\in\Sigma}}
\!\!\!\!
\int_\k
\big[{\colorb a_\k}\,{\mathbbm T}\big]_\u 
\big[{\colorb a_\k^*}\,{\mathbbm T}\big]_\v
+\int\limits_{_{\vec\u\in\Sigma}}
\big[{\colorb{\bs\alpha}}\,{\mathbbm T}\big]_\u
\Bigg]\;
\left.{\colorb\frac{dN_g}{dyd^2\vec\p_\perp}}\right|_{_{\rm LO}}\; ,
\label{eq:LONLO}
\end{equation}
where ${\bs\alpha}$ and $a_\k$ are small perturbations to the
classical field ${\cal A}$, $\Sigma$ a surface used to set the initial
condition for the fields (e.g. the light-cone), and ${\mathbbm T}_\u$
the functional derivative with respect to the initial classical field
at the point $\u\in\Sigma$. In this formula, the integration over $\k$
has a logarithmic dependence on the cutoff
\begin{eqnarray}
    \frac{1}{2}\!\!\!\!
\int\limits_{_{\vec\u,\vec\v\in\Sigma}}
\!\!\!\!
\int_\k
\big[{\colorb a_\k}\,{\mathbbm T}\big]_\u 
\big[{\colorb a_\k^*}\,{\mathbbm T}\big]_\v
+\int\limits_{_{\vec\u\in\Sigma}}
\big[{\colorb{\bs\alpha}}\,{\mathbbm T}\big]_\u
=
    \ln\left(\Lambda^+\right)\,{\colorb{\cal H}_1}
    +
    \ln\left(\Lambda^-\right)\,{\colorb{\cal H}_2}
    +
\cdots
\end{eqnarray}
where ${\cal H}_{1,2}$ are the JIMWLK Hamiltonians of the two nuclei.
This formula is the formal realization of the causality argument
exposed at the beginning of this section, since the logarithms
associated to the two cutoffs are accompanied by the JIMWLK of the
corresponding nucleus, without any mixing. Thanks to this formula, one
can absorb the logarithms of the cutoff in JIMWLK-evolved
distributions of color sources,
\begin{equation}
\left<\frac{dN_g}{dy d^2\vec\p_\perp}\right>_{_{\rm Leading Log}}
=
\int 
\big[D{\colora\rho_{_1}}\,D{\colorb\rho_{_2}}\big]
\;
{\colora W_1\big[\rho_{_1}\big]}\;
{\colorb W_2\big[\rho_{_2}\big]}
\;
{\frac{dN_g[\rho_{1,2}]}{dy d^2\vec\p_\perp}}\; .
\end{equation}
The same factorization holds for multi-gluon inclusive
spectra\cite{GelisLV4,GelisLV5}, and more generally for all the
inclusive observables, for which eq.~(\ref{eq:LONLO}) is valid.

\section{Extension to quark production (work in progress)}
A natural extension of these results is to study it in the case of
quark production. To that effect, one needs expressions for the quark
spectrum at LO and NLO (see the two diagrams on the right of the
figure \ref{fig:gluon}). These contributions have been worked out, and
it turns out that they are related by a formula similar to
eq.~(\ref{eq:LONLO})\cite{GelisL1},
\begin{eqnarray}
&&
\left.{\colorb\frac{dN_{_Q}}{dyd^2\vec\p_\perp}}\right|_{_{\rm NLO}}
=
\smash{
\Bigg[
 \frac{1}{2}\!\!\!\!
\int\limits_{_{\vec\u,\vec\v\in\Sigma}}
\!\!\!\!
\int_\k
\big[{\colorb a_\k}\,{\mathbbm T}+\underline{{\colorb b_\k}\,{\mathbbm T}_{_\psi}}\big]_\u 
\big[{\colorb a_\k^*}\,{\mathbbm T}+\underline{{\colorb b_\k^\dagger}\,{\mathbbm T}_{_\psi}}\big]_\v}
\nonumber\\
&&\qquad\qquad\qquad\qquad\quad
{
+\int\limits_{_{\vec\u\in\Sigma}}
\big[{\colorb{\bs\alpha}}\,{\mathbbm T}+\underline{{\colorb{\bs\beta}}\,{\mathbbm T}_{_\psi}}\big]_\u
\Bigg]\;
\left.{\colorb\frac{dN_{_Q}}{dyd^2\vec\p_\perp}}\right|_{_{\rm LO}}}\; ,
\end{eqnarray}
where the underlined terms are new and contain purely fermionic
quantities~: $b_\k$ and ${\bs\beta}$ are small fermionic fields,
and where the operator ${\mathbbm T}_\psi$ is a functional derivative
with respect to the initial value of fermion fields on $\Sigma$. This
formula is only the first step; the second part of this program
consists in extracting the logarithms from the new fermionic terms in
the operator that appears in the right hand side, and to prove that
they are also proportional to the JIMWLK Hamiltonian.

\section*{Acknowledgements}
This work is supported by the ANR project \#~11-BS04-015-01.

%\bibliographystyle{unsrt}
%\bibliography{biblio}

\end{document}